# The Climate of the Khagan. Observations on palaeo-environmental Factors of the History of the Avars (6th-9th century AD)*

Johannes Preiser-Kapeller, OEAW (Vienna)

**Abstract**

Based on palaeoenvironmental, historical and archaeological data, the paper proposes possible climatic impacts on the history of the Avar Khaganate, which comprised the Carpathian Basin between the late 6th and the early 9th century AD. While the establishment of the Avars in East Central Europe took place within a period characterised by cold and dry climatic conditions (recently identified as "Late Antique Little Ice Age"), more stable climatic parameters may have favoured the stabilisation of Avar rule after a crisis in the aftermath of 626 AD. Data indicates growth of settlement and agricultural activity up to the mid-8th century. These developments did not necessarily strengthen central power, but may have contributed to a greater autonomy of various groups on the basis of increased resources. The Khaganate quickly disintegrated faced by the Carolingian advance of the 790s; the last decades of documented Avar presence were again accompanied by environmental vicissitudes.

In an article published in 2016, Büntgen et alii on the basis of palaeoenvironmental data from Central Europe and Central Asia identified a prolonged cool period across Eurasia between 536 and 660 AD, initiated by a series of volcanic eruptions, which they described as "Late Antique Little Ice Age" (= LALIA) (**fig. 1,** top). Among the various political and socio-economic upheavals during these decades between the Mediterranean and China, the authors listed the advance of the Avars to the Danube and the following migration movements, especially of Slavs, towards the Byzantine Balkans[1]. Recently, also McCormick and colleagues as well as Cook have mentioned adverse climatic conditions in the Central Asian steppes as one factor contributing to the collapse of the Rouran Khaganate in ca. 552 AD and the following movement of groups, which we encounter as Avars in Byzantine and other sources, to the Carpathian Basin; Cook in tree ring data from Dulan-Wulan in North-Central China identified a "multi-decadal megadrought" in the mid-6th century[2]. And most recently, Sümegi et alii have addressed the question of climate change in the Avar period on the basis of a regional study on Lake Baláta in southwest Hungary[3]. On the basis of these new findings, the present volume devoted to one of the leading experts in the archaeology of the Avars provides the opportunity to further reflect on possible climatic factors to the history of this people and their Khaganate, which comprised the Carpathian Basin and adjacent areas between the 6th and the 9th century AD[4].

---

* Research for this paper was undertaken within the framework of the Wittgenstein Prize-project of Prof. Claudia Rapp (Vienna) "Moving Byzantium: Mobility, Microstructures and Personal Agency" (http://rapp.univie.ac.at/), of which Prof. Falko Daim is an affiliate scholar. I would also like to thank Prof. Ewald Kislinger (Vienna) and Prof. Pál Sümegi (Szeged) for most valuable comments and amendments to this paper.

[1] Büntgen et alii, Cooling and societal change. Cf. Gunn (ed.), Year without Summer; Stathakopoulos, Famine nr. 92 (on the "dust-veil event" of 536); McCormick et alii, Climate Change; Haldon et alii, Climate. For the historical events cf. Pohl, Awaren 29-37.

[2] McCormick et alii, Climate Change; Cook, Megadroughts.

For this purpose, we use the temperature reconstructions provided by Büntgen et alii in their study, who integrated tree ring data from the European Alps and the Russian Altai in order to model climatic fluctuations across the main theatre of Avar history from the Danube to Central Asia (**fig. 1,** top)[5]. In addition, Büntgen and colleagues in 2011 created a reconstruction of spring precipitation in Central Europe for the last two Millennia, which supplies a further palaeo-environmental parameter of Avar presence in Pannonia on a wider scale (**fig. 1,** middle and bottom)[6]**.** Based on similar data, Cook et alii present spatial reconstructions of summer wetness and dryness across Europe and the Mediterranean for the last 2000 years in their "Old World Drought Atlas" (OWDA); for the period under consideration, the OWDA fortunately also covers the Carpathian Basin[7]. In contrast to these annually resolved over-regional reconstructions, other proxy data from the Carpathian Basin such as lake sediments and pollen provide more locally confined information on changes in the landscape on a long term scale, caused both by changing climate conditions and/or human agency (**fig. 2**)[8]**.** These "archives of nature" can be compared with historical[9] and archaeological data as "archives of society"; also here, over-regional, regional and local as well as centennial, decadal and annual scales can be observed[10].

Especially extreme events such as droughts, colds or floods affected human societies; they damaged vegetation (pastures, but also arable crops of agricultural communities within nomadic empires) and could lead to a reduction of water resources (especially in the semiarid areas of Central Asia). This in turn affected livestock, which also could be exposed to an increased risk of epizootics due to specific weather conditions. The same was true for humans (e. g. the outbreak of the "Justinianic plague" during LALIA[11]), who equally suffered from malnutrition. Hunger would first affect younger and older as well

---

[3] Sümegi et alii, Extreme Dry Climate.
[4] From the numerous studies of Falko Daim we would like to mention especially: Daim (ed.), Awarenforschungen; Daim et alii (eds.), Hunnen und Awaren; Daim, Structures of Identification; Daim (ed.), Die Awaren am Rand; Daim, Avars and Avar Archaeology.
[5] Büntgen et alii, Cooling and societal change.
[6] Büntgen et alii, Climate Variability. Another relevant recent (oak) tree ring series ranging back to 761 AD stems from the Czech Republic: Dobrovolný et alii, A tree-ring perspective.
[7] Cook et alii, Old World megadroughts. The Balkans and the Eastern Mediterranean unfortunately are not covered during the 6th to 9th centuries.
[8] See the various types of data (and also the problems with their dating and use) mentioned below. Cf. Bojňanský / Fargašová, Atlas of Seeds and Fruits; Werger / van Staalduinen (eds.), Eurasian Steppes 209-252.
[9] For historical sources from the neighbouring Byzantine Balkans as well as Merovingian/Carolingian Central Europe, we possess now a number of exhaustive catalogues which are used and cited for the present study: Telelis, Μετεωρολογικά φαινόμενα; Stathakopoulos, Famine; Newfield, The contours of disease and hunger; McCormick et alii, Geodatabase (in the studies of Telelis and Newfield also full citations of sources can be found).
[10] For a longer overview on these types of data, their scales and the problems of their temporal and spatial resolutions cf. Preiser-Kapeller, Collapse (with further literature). Cf. also Juhász, Reconstitution palynologique; Zatykó / Juhász / Sümegi, Environmental archaeology; Izdebski et alii, Palynological Data.
[11] Stathakopoulos, Famine, esp. nr. 159 for an outbreak of the plague among the Avars reported for the year 598 during a campaign in Byzantine Thrace (while this event was recorded in Byzantine sources, we learn nothing about outbreaks of the plague in the Avar core regions in the Carpathian Basin). For the plague as a possible factor

as poorer members of a community, but also delegitimise the claim to leadership of elites and increase social conflict. Especially in the case of nomads, such crises could motivate mobility in search for other sources of resources, thus leading to conflict with neighbouring groups, both nomadic and sedentary[12]. Besides these general impacts, after the establishment of Avar rule in the Carpathian Basin **(fig. 2)** one has to take in into consideration specific natural hazards in these areas, which included both extreme colds and droughts, as documented in fluctuating water levels of Lake Balaton and Lake Fertő/Neusiedlersee, for instance[13]. On the other hand, before the hydro-engineering measures of the 19th and 20th century wide flood plains characterised the areas along Danube, Tisza and their tributaries (summing up to 12-15 % of the entire Carpathian Basin, ca. 30,000 km²). Extreme flood events resulted from increased precipitation and snowmelt, especially in the upper (alpine) catchment of the Danube, or extreme cold, leading to devastating ice (jam) floods[14]. Resulting limitations to permanent settlement are well documented for the high and late Middle Ages; recent studies indicate similar environmental factors for settlement site selection at the edges of floods areas in the Avar period[15]. Floods would equally impede military movements, also of nomadic troops, as a recent study of Büntgen and Di Cosmo on the Mongol invasion of 1242 illustrates[16]. Landscape degradation in the Carpathian Basin, in turn, has been attributed to nomadic pastoralism of Huns, Avar, Magyars and Cumans between the 5th and 13th century, a still ongoing debate[17]. Without doubt, Avar pastoralism created a huge demand of pasture lands for economic and military purposes (for an army of maybe 20,000 horsemen)[18]. Yet their realm always and increasingly included permanent settlements and depended on agriculture, as also archaeobotanical evidence illustrates[19].

---

contributing to Avar and Slavic expansion on the Byzantine Balkans and its potential effect on pastoral societies cf. also Sołtysiak, The plague pandemic. For a more nuanced discussion of the potential effect of plague epidemics on nomadic communities and the debates on this issue cf. also Varlik, Plague and Empire 113-118, and Franz / Riha / Schubert, Plagues in Nomadic Contexts. Remains of the black rat (*rattus rattus*) have been found in Hungary within and beyond the Roman borders, dating to the 3rd-4th century; unfortunately, there is no archaeological evidence for the following period until the 13th century, cf. Kovács, Dispersal History.

[12] For a detailed analytical framework for the impact of environmental hazards on pre-modern societies cf. Krämer, Hungerkrise. Cf. also Newfield, The contours of disease and hunger, esp. for epizootics, and Telelis, Environmental History.

[13] Vadas / Rácz. Climatic Changes; Eitzinger et alii, Auswirkungen einer Klimaänderung; Soja et alii, Climate impacts; Sümegi et alii, Lake Balaton; Kiss, Floods 65-73; Lóczy (ed.), Landscapes 19-24.

[14] Kiss, Floods, esp. 21-38 (with detailed discussion of the hydrology of the area); Brilly (ed.), Hydrological Processes, esp. 25-77 and 121-123; Lóczy (ed.), Landscapes 24-27; Vadas / Rácz. Climatic Changes; Sümegi et alii, Long environment change; Sümegi, Link.

[15] Bugarski, The Geomorphological Matrix; Odler, Avarské sídliská; Pinke et alii, Settlement Patterns; Sümegi, Link.

[16] Büntgen / Di Cosmo, Climatic and environmental aspects.

[17] Werger / van Staalduinen (eds.), Eurasian Steppes 209-252 (highlighting the most significant modifications of landscapes during the last 200 years); Sümegi et alii, Long environment change, esp. 19-20; Sümegi, Link. See also Daim, Avars and Avar Archaeology 485.

[18] Pohl, Awaren 37-38, 189-191; Curta, Avar Blitzkrieg, esp. 73-75; Priskin, Horses. Cf. also Bowlus, Lechfeld 22-27, on the "carrying capacity" of the Carpathian Basin for steppe nomadic pastoralism and warfare.

[19] Pohl, Awaren 191-193; Curta, Southeastern Europe 65-66; Vida, Raumkonzepte; Gyulai, Archaeobotany in Hungary; Gyulai, The History of Broomcorn Millet; Rapan Papeša / Kenéz / Pető, The Archaeobotanical Assessment; Noche-Dowdy, Multi-Isotope Analysis. Cf. Daim, Structures of Identification 83-84. Cf. also Hämäläinen, Comanche Empire, esp. 351-353, and Hämäläinen, Kinetic Empire, on the dependence of pastoral federations on "viable agricultural societies" across and within their borders. Hämäläinen describes a "kinetic

**The Early Avar period (568 to 630 AD)**[20]

During LALIA, a first period of extremely cold summers affected Central Europe from 535 to 558; spring precipitation was low from the late 530s to the early 570s, with minima in 544 and 567 (**fig. 1, middle**). Also written sources register various weather extremes in Central and Western Europe during these decades[21]. Byzantine texts report an extreme winter for 558/559, which allowed "Barbarian" groups to cross the frozen Danube[22]. We also have more specific proxy data for the Carpathian Basin and neighbouring regions such as the Upper Dniester valley, where a transition to a period of increased floods took place before the last quarter of the 6th century[23]. Pollen, macrofossil and sediment analyses at Lake Nádas in northern Hungary indicate a severe drop of temperatures in the 6th century and rapid change from very low water levels in the 5th century to high levels in the 6th and 7th centuries, accompanied by increased annual precipitation (**fig. 3**)[24]. For other regions, however, a change towards drier conditions can be observed in the 6th century, as for the Tăul Muced bog in the Eastern Carpathian Mountains; for Lake Balaton, in turn, recently rising water levels have been reconstructed for the 5th to 7th century[25]. Thus, we have to reckon with different regional shapes of the general trend of LALIA; the overall impact on agricultural communities, however, seems to have been negative. Trajectories of grain pollen for three sites from Hungary, where sample size was sufficient of our period of interest (Lake Balaton, Nagy-Mohos and Pölöske), all show a downwards trend in the 6th century (**fig. 4**, with citations of data sources); but the dating for these layers is highly problematic in all three sites, which very much limits (or maybe even invalidates) their explanatory value for the Avar period.[26] Yet at least similar trajectories can be observed in a recent synthetic (and significantly more reliably datable) reconstruction of agricultural outputs for two neighbouring areas, the region south of Cracow (Poland) and Bohemia, undertaken by Izdebski et alii[27].

The first period of extreme cold was followed by an interval of shorter temperature fluctuations from 560 until the early 590s (**fig. 1**, middle); in these decades between 568 and 582 (conquest of Sirmium)

---

empire" as "a flexible imperial organization that revolves around a set of mobile activities and relies on selective nodal control of key resources". On spatial concepts of the Avars cf. Vida, Raumkonzepte.

[20] This papers follows the new chronological division proposed in Stadler, Quantitative Studien, see also Stadler, Avar Chronology; on the older chronology see Daim, Avars and Avar Archaeology; Breuer, Byzanz an der Donau. For the historical events of this period cf. Pohl, Awaren 27-162, 237-255; Curta, Southeastern Europe 61-69; Kardaras, Ἄβαροι 37-136.

[21] McCormick et alii, Geodatabase nr. 540-584; McCormick et alii, Climate Change. See also Cook et alii, Old World megadroughts.

[22] Telelis, Μετεωρολογικά φαινόμενα nr. 176, 254-255; Kislinger, Angriff; Pohl, Awaren 21; Curta, Avar Blitzkrieg 77.

[23] Gębica et alii, Medieval accumulation.

[24] Sümegi et al., Middle Age, esp. 280-285 (with table 6), 289-290; Vadas / Rácz, Climatic Changes 212 (fig. 2); Kiss, Floods 61-62 (with fig. 11). Comparative data stems from the Nyírjes-tó peat bog of Sirok in northern Hungary, cf. Náfrádi et alii, Future Climate Impacts.

[25] Gałka et alii, A 9000 year record. Kiss, Floods 65-67, sums up earlier reconstructions of the water level of Lake Balaton, but stresses their problematic character. Cf. also Vadas / Rácz, Climatic Changes 206-207. For the most recent reconstructions of Lake Balaton water levels for the 5th-9th century, cf. Sümegi et alii, Reconstruction.

[26] I would like to thank Prof. Pál Sümegi (Szeged) for drawing my attention to the highly problematic character of this data.

[27] Izdebski et alii, Palynological Data, esp. 17-24 and 27. On connections of the Cracow region to the Avars cf. Rudnicki, New Avar finds.

the establishment of Avar power in the Carpathian Basin took place[28]. Besides climatic vicissitudes, violence, unrest and migrations (such as the emigration of part of the Lombards to Italy or the resettlement of Gepids) accompanying this process damaged existing agricultural communities; recent studies, however, indicate also a significant continuity of settlement and of networks of exchange within and beyond the Carpathian Basin, into which the newcomers now (forcefully) integrated[29].

A second prolonged cold period occurred between 595 and 615, while drier conditions continued between 604 and 614 and between 617 and 625 (**fig. 1**, middle). These years were characterised by frequent warfare between the Avars and the Byzantine Empire. From written sources, we especially learn about impacts of extreme weather on the Byzantines and their troops; for 599, they register a cold winter which also led to the death of draft animals[30]. For 602, we read about a famine in Constantinople early that year and colds and rains affecting the army in the Balkans in autumn, both contributing to the fall of Emperor Maurice[31]. A drought year is referred to for Constantinople in 610, when Emperor Phokas was replaced by Heraclius[32]. For the very dry year of 618, sources report another famine in Constantinople (also caused by the loss of Egypt to the Sasanians)[33]. For the early Avar realm, success in these wars was essential; as Falko Daim has explained: "if it were not for Byzantium, the Avar Empire would never have been founded, because large annual payments from Constantinople and rich loot from raids on the Balkans stabilised the khagan's power"[34]. Accordingly, the successful Byzantine campaigns of the 590s brought about a first crisis of the Khaganate, which was saved by the violent end of Maurice and the following internal turmoil in Byzantium. A second crisis emerged after the failure of the Avar siege of Constantinople in 626, in the midst of another cold period with a minimum in 626/627 (**fig. 1**, middle)[35]. The following years saw a civil war in the Khaganate, the rise of the realm of Samo in Bohemia and the creation of "Great Bulgaria" under Kuvrat in the Ukrainian steppes[36]. These perturbations at ca. 630 also led to the end of the important settlement of Keszthely-Fenékpuszta, a former Roman fortress at Lake Balaton, where a "sub-Mediterranean" agriculture based on the cultivation of wine, wheat and walnut had continued until the 7th century, as recent palaeo-environmental studies by Sümegi et alii have demonstrated **(fig. 2)**[37].

---

[28] Pohl, Awaren 52-76; Kislinger, Regionalgeschichte 25-40, 72-101.
[29] Pohl, Awaren 225-236; Curta, Southeastern Europe 61-69; Vida, Conflict and coexistence; Vida, Local and Foreign Romans; Stadler, Quantitative Studien, esp. 130-133; Stadler, Ethnische Gruppen; Koncz, 568. Cf. also Daim, Avars and Avar Archaeology, esp. 469-476.
[30] Telelis, Μετεωρολογικά φαινόμενα nr. 203, 275-276. On the Avar-Byzantine wars cf. Pohl, Awaren 128-162, 237-256; Curta, Avar Blitzkrieg; Kardaras, Ἄβαροι 70-136; Hurbanič, Vpády Avarov. Cf. Bowlus, Lechfeld 27-36, on possible limits of humid weather for nomadic warfare, especially for the use of composite bows.
[31] Telelis, Μετεωρολογικά φαινόμενα nr. 205, 277-279 and nr. 206, 279-280; Stathakopoulos, Famine nr. 165; Pfeilschifter, Kaiser 252-293. Cf. also the reconstruction of Cook et alii, Old World megadroughts.
[32] Telelis, Μετεωρολογικά φαινόμενα nr. 212, 288-289.
[33] Telelis, Μετεωρολογικά φαινόμενα nr. 215, 291-292; Stathakopoulos, Famine nr. 173. Cf. Cook et alii, Old World megadroughts, who reconstruct a severe drought across large parts of East Central and Eastern Europe for 618.
[34] Daim, Byzantine Belt Ornaments 62. Cf. Whitby, Emperor Maurice (for the Byzantine-Avar wars in the 590s); Pohl, Awaren 178-185, 205-215; Curta, Avar Blitzkrieg; Curta, Southeastern Europe 64-65; Breuer, Byzanz an der Donau; Csiky / Magyar-Harsheghyi, Wine. Cf. also Hämäläinen, Comanche Empire, esp. 349-350, and Hämäläinen, Kinetic Empire, for "expansionist and exploitative" strategies of nomadic empires with regard to the material wealth and also labour force of neighbouring societies. For similar observation on the Avars see also Vida, Raumkonzepte.
[35] Pohl, Awaren 156-162; Curta, Southeastern Europe 68-69, 75-76; Kardaras, Ἄβαροι 86-102.
[36] Daim, Avars and Avar Archaeology 481-482; Pohl, Awaren 256-282; Curta, Southeastern Europe 76-78; Kardaras, Ἄβαροι 118-126, 137-155. On the siege of 626 cf. also Hurbanič, Posledná.

**The Middle Avar period (630-680 AD)**

The crisis in the aftermath of 626 was followed by a phase of "re-organisation of the Avar Empire" between the 630s and 670s, as Daim has outlined; from the mid-7th century onwards one observes also again an increased inflow of Byzantine coins and objects, now presumably through trade and as diplomatic gifts rather than from plunder and as tribute. Towards the end of this period, the "process during which the Avar Empire had gradually re-gained its strength had (…) been more or less completed"[38]. This process included also the extension of settlement in various areas, such as the Vojvodina, along the Danube to the west and in the core region of Avar power between Danube and Tisza **(fig. 2)**[39]. Confirming these archaeological observations, data from Lake Baláta **(fig. 2)** indicates increasing stock-breeding activity as well as cereal cultivation from the mid-7th century onwards.[40] Much less reliable (see above), also the trajectories of grain pollen from three sites in Hungary show an upwards trend of agricultural output in the 7th century **(fig. 4)**. Also Curta assumes population growth during this period, contributing and benefiting from socio-economic changes, which also have been proposed by Herold on the basis of her re-interpretation of Avar ceramics. For her the emergence of "slow-wheel-turned pottery" in the Middle Avar period "reflects a change in the organisation of pottery production, which was most likely connected to other (economic) changes in the Avar Khaganate that are less readily detectable by archaeological methods"[41].

Consolidation and growth would also have been supported by climatic trends; the reconstructions of Büntgen et alii still show often fluctuating temperature and precipitation conditions for these decades, but not long periods of more extreme colds and droughts as in the preceding period **(fig. 1**, middle**)**. Some extremely dry years (647, 662 and 671) and a cold one (648) clustered around the middle of the 7th century; for 647/648 we also learn about heavy storms in Constantinople[42]. From the Carpathian Basin, the data of Lake Nádas indicates an increase of temperatures from the mid-7th century onwards **(fig. 3)**[43].

---

[37] Daim, Avars and Avar Archaeology 473-476; Pohl, Awaren 282; Sümegi et alii, Reconstruction, esp. 557-564; Sümegi et alii, Fenékpuszta, esp. 11-12; Varga et alii, Multivariate Statistics. For the import of wine from the Byzantine Mediterranean cf. Csiky / Magyar-Harsheghyi, Wine.
[38] Daim, Avars and Avar Archaeology 481-490, 517; Daim, Byzantine Belt Ornaments 66-67; Pohl, Awaren 282-287; Curta, Southeastern Europe 90-92.
[39] Stadler, Ethnische Gruppen 126-127 (who assumes that the expansion to the west also included the "re-integration" of Slavic groups temporarily part of the realm of Samo between 630 and 660); Balogh, Problems; Bugarski, The Geomorphological Matrix; Odler, Avarské sídliská; Noche-Dowdy, Multi-Isotope Analysis 25-26; Knipl / Sümegi, Life at the interface, esp. 445; Curta, Southeastern Europe 92.
[40] Sümegi et alii, Extreme Dry Climate, esp. 481-484.
[41] Curta, Southeastern Europe 92-93; Herold, Insights 225-227.
[42] Telelis, Μετεωρολογικά φαινόμενα nr. 224, 299-302. Cf. Cook et alii, Old World megadroughts.
[43] Sümegi et alii, Middle Age; Vadas / Rácz, Climatic Changes 212 (fig. 2).

**The Late Avar period and the end of Avar rule (680-822 AD)**

Agricultural and demographic growth did not necessarily strengthen the central power, but may have contributed to a greater autonomy of various groups and regions within the Khaganate on the basis of increased resources[44]. Already the turn from the middle to the lave Avar period was marked in 680/681 by a conflict between the Khagan and a group of descendants of deportees from the Balkans under the leadership of Kouber; obviously, they possessed a distinct organisation as community, left the Avar realm and crossed the Danube until reaching Byzantine territory[45].

Also climatic condition became more turbulent again, with a dramatic cooling trend between 685 and 690, the coldest summer in the late period; strong ups and downs also characterised precipitation, with minima in 688 and especially 695, one of the driest years in this period, while 700 was extremely wet (**fig. 1**, middle).[46] Warmer conditions returned from 695 to 709, followed by another short term cooling with a minimum in 714 (**fig. 1**, middle and bottom). Accordingly, we learn about harsh winters and great floods in Carolingian sources for 709-711, floods in Rome in the winter of 716/717 and an extreme winter during the Arab siege of Constantinople in 717/718[47]. Such extreme colds were not seen again until the mid-740s (**fig. 1**, bottom).

Archaeological evidence and data from Lake Baláta (**fig. 2**)[48] suggest a continuation of growth until the mid-8th century; this may have led to a "better organised and more 'egalitarian' society for the Late Avar Period, also reflected in the metal finds, which become much more widespread and more unified in the Late Avar Period"[49]. This could have further promoted a "shift of power away from the centre", as Daim has called it, with border regions gaining greater weight and the emergence of regional centres such as Carantania at the western periphery of the Khaganate, which from the 740s onwards allied with Bavaria[50]. Also Peter Stadler has observed regional differentiation on the basis of material findings and identifies 14 regional clusters, which were characterised by more dense internal exchange[51]. Between the middle and end of the 8th century, however, growth of settlement and agricultural output seems to have been replaced by stagnation or even recession in some regions of the Khaganate; archaeologists state an "end of colonisation" between Danube and Tisza and of the expansion of settlement in modern-day Slovakia around that time, for instance[52]. The (problematic) pollen data from the three sites already mentioned (**fig. 4**), but especially also the analysis of the much better dated sediments in Lake Baláta in

---

[44] Cf. Hämäläinen, Comanche Empire, esp. 348-350, and Hämäläinen, Kinetic Empire, for similar observations on the interplay between economic expansion, centrifugal forces and cohesion of a nomadic polity.
[45] Pohl, Awaren 278-281; Curta, Southeastern Europe 106-107; Kardaras, Ἄβαροι 177-179.
[46] Also confirmed in the reconstruction of Cook et alii, Old World megadroughts.
[47] McCormick et alii, Geodatabase nr. 687 and 688; Telelis, Μετεωρολογικά φαινόμενα nr. 249, 321-323; Stathakopoulos, Famine nr. 208; Newfield, The contours of disease and hunger 418 (nr. 1-11); Glaser, Klimageschichte 56-57.
[48] Sümegi et alii, Extreme Dry Climate, esp. 481-484.
[49] Herold, Insights 225-227; cf. Curta, Southeastern Europe 92-93.
[50] Daim, Avars and Avar Archaeology 503-504, 511; Pohl, Awaren 292-308; Curta, Southeastern Europe 92-95.
[51] Stadler, Quantitative Studien, esp. map 201; Stadler, Ethnische Gruppen, esp. 118-120, 126-127.
[52] Balogh, Problems; Odler, Avarské sídliská.

Southwestern Hungary indicate downwards trends of agricultural output in the second half of the 8th century[53].

Again, less beneficial climatic conditions may have contributed to this reversal of trends; locally, the data from Lake Nádas shows a change towards drier conditions from the 7th to the 8th century (**fig. 3**)[54]. In the data of Büntgen et alii, drier spring conditions show up between 747 and 750 with 749 as extremely dry year (**fig. 1**, bottom). Not visible in their summer temperature reconstruction is the extreme winter of 763/764, which was probably caused by another major volcanic eruption and is registered both in Latin and in Byzantine sources from Western Europe to the Balkans; but it shows up as a negative extreme in the tree ring series from the Czech Republic for the year 764[55]. A short term cold occurred from 771 to 775 (when we learn about heavy storms in Thrace in 774[56]), and two further cold spells until 785/786 (when sources register a harsh winter in Saxony in 785 followed by floods impeding the campaign of Charlemagne) (**fig. 1**, bottom)[57]. Stronger fluctuations also continued in precipitation until the end of the 8th century, with dry spells around 764, 773, 782, 790 and especially 797 (**fig. 1**, bottom). The reconstruction is confirmed by news about droughts in Constantinople in 764 and 766/767, in Saxony in 772, in the Moselle region in 783 and in Burgundy in 797[58]. A wet year on the contrast is documented both in the precipitation reconstruction and in the written sources for 793, when constant rain impeded the digging of a canal between Rhine and Danube as initiated by Charlemagne[59].

This canal may also have played a role in the strategic planning of Charlemagne against the Avar Khaganate, which he first attacked in 791. Initial successes were however undone by an equine epizootic, which damaged the war horses of the Carolingian army. This disease may as well have affected the Avars; as Newfield points out: "(…) it is possible that the Avars sustained considerable losses in Pannonia and that this may have partially accounted for their military decline and defeat in 796". The same may have been true for two epizootics registered in Carolingian sources in 801 and especially in 809/810, when it is said that "Noricum (…) suffered these things in particular, together with the neighbouring regions to it". This epidemic was accompanied by a widespread drought across Central and Western Europe, as also reflected in the reconstruction of the OWDA (**fig. 5**) and in the tree ring series from the Czech Republic for 810[60]. At this time, the Khaganate after civil wars following the campaign of 791 and a Frankish advance towards the central "Ring" in 796 had already disintegrated into various principalities, many of which were now nominally under Carolingian suzerainty[61]. The last reference to an Avar political entity is dated to 822 AD[62].

---

[53] Náfrádi et alii, Reconstruction 1567, 1571-1573; Sümegi et alii, Extreme Dry Climate, esp. 489-490.
[54] Kiss, Floods 61-62 (with fig. 11).
[55] McCormick et alii, Geodatabase nr. 743; Telelis, Μετεωρολογικά φαινόμενα nr. 271, 342-350; Newfield, The contours of disease and hunger 308-310, 418-419 (nr. 18-33); McCormick / Dutton / Mayewski, Volcanoes and the Climate Forcing, esp. 878-881; Dobrovolný et alii, A tree-ring perspective (supplementary material); Glaser, Klimageschichte 56-57.
[56] Telelis, Μετεωρολογικά φαινόμενα nr. 278, 357.
[57] McCormick et alii, Geodatabase nr. 757; Newfield, The contours of disease and hunger 422 (nr. 45).
[58] Telelis, Μετεωρολογικά φαινόμενα nr. 272, 351-352 and nr. 275, 355-356; McCormick et alii, Geodatabase nr. 744, 745, 755, 768; Newfield, The contours of disease and hunger 420-421 (nr. 35, 42), 428 (nr. 66). Cf. Cook et alii, Old World megadroughts.
[59] McCormick et alii, Geodatabase nr. 765; Newfield, The contours of disease and hunger 78, 427 (nr. 62); Ettel et alii (eds.), Großbaustelle. The spatial reconstruction of Cook et alii, Old World megadroughts, suggests that these extremely moist conditions were confined to the region around the canal, while droughts affected other parts of the Carolingian Empire.

At this time, another longer cold period (from 821 to 840) had started (**fig. 1**, bottom); Byzantine sources report strong colds, heavy storms and famines in the early 820s[63]. Diseases affected Carolingian troops in "Upper Pannonia" at the river Drave in 820, while equally strong rain, floods and bad harvests afflicted the Carolingian lands in 820-822[64]. Again, volcanic eruptions possibly contributed to these weather extremes across Europe[65]. Thus the Avar era ended at it had begun – with a clustering of extreme weather events.

**Conclusion**

Even before the recent studies mentioned at the beginning of this paper, climatic factors have been proposed as motivator of the migration of the Avars and the emergence of their Khaganate[66]. The same is true for its demise; Györffy and Zólyomi for instance in 1994 hypothesised "that a relatively dry climate was predominant in the Carpathian Basin until the mid-8th century, and probably was decisive in the demise of the Avar Empire in the Carpathian Basin", since it reduced the resource basis for nomadic pastoralism[67]. As also our short study on the basis of most recent data has shown, "there were almost certainly environmental as well as political factors in the break-up of the Avar Empire"[68]. But when comparing palaeo-environmental, historical and archaeological data, "one must exercise caution in interpreting the correlations of environmental and social crises"[69]. Simple, mono-causal scenarios cannot do justice to the complex interplay between socio-economic, political and environmental dynamics, especially in their regional variations[70]. In contrast to the hypothesis of Györffy and Zólyomi[71], our reconstruction indicates that it may have been the growth of settlement and resources, favoured by more stable climatic conditions after the end of the Late Antique Little Ice Age, which

---

[60] Newfield, The contours of disease and hunger 76-78 (with fn. 242), 169-170, 196-203 (for epizootics in 801 and 809/810), 423-424 (nr. 50-51), 432-435 (nr. 86-91, with translations of sources); Newfield, A great Carolingian panzootic, esp. 200-204; Cook et alii, Old World megadroughts; Dobrovolný et alii, A tree-ring perspective (supplementary material).

[61] Cf. Pohl, Awaren 312-328; Curta, Southeastern Europe 130-131.

[62] Pohl, Awaren 323.

[63] Telelis, Μετεωρολογικά φαινόμενα nr. 302, 379-381, nr. 303, 381-382, nr. 307, 384-385 and nr. 308, 385.

[64] Newfield, The contours of disease and hunger 437-440 (nr. 100-110); Glaser, Klimageschichte 56-57. Cf. the OWDA-reconstruction of Cook et alii, Old World megadroughts.

[65] McCormick / Dutton / Mayewski, Volcanoes and the Climate Forcing 881-884.

[66] For an overview of such approaches see Rácz, Magyarország környezettörténete; Vadas / Rácz. Climatic Changes.

[67] Györffy / Zólyomi, A Kárpát-medence és Etelköz; Vadas / Rácz. Climatic Changes 208-209; Rapan Papeša / Kenéz / Pető, The Archaeobotanical Assessment 265-266.

[68] Vadas / Rácz. Climatic Changes 210.

[69] Sümegi et alii, Middle Age, esp. 295 for the citation; Vadas / Rácz, Climatic Changes 212.

[70] Cf. also Preiser-Kapeller, Collapse.

[71] See also Sümegi et alii, Reconstruction, esp. 563 (fn. 85), and Sümegi et alii, Extreme Dry Climate, on this scenario and its problems.

contributed to a differentiation and decentralisation of power within the Khaganate even before the Carolingian offensive. The reversal of this trend at the mid-8th century as outlined above, however, cannot be observed in the neighbouring regions of Bohemia and Southern Poland, where pollen data indicates on the contrast the beginning of agricultural growth in this period[72]. And while the disintegration of the Avar Empire at the turn from the 8th to the 9th century was once more accompanied by a series of natural disasters, similar conditions affected the Carolingian realm without restricting its ability to expand its power into the Carpathian Basin. Climatic parameters thus were relevant, but on their own not sufficient to explain that the "Avar Empire did not come to grips with the transition to a medieval state"[73]. Further data and research is necessary to better estimate their actual contribution.

---

[72] Izdebski et alii, Palynological Data, esp. 17-24 and 27.
[73] Daim, Avars and Avar Archaeology 523.

Johannes Preiser-Kapeller

Österreichische Akademie der Wissenschaften

Institut für Mittelalterforschung/Abteilung Byzanzforschung

Hollandstraße 11-13/4

A-1020 Wien

Johannes.Preiser-Kapeller@oeaw.ac.at

**Figures**

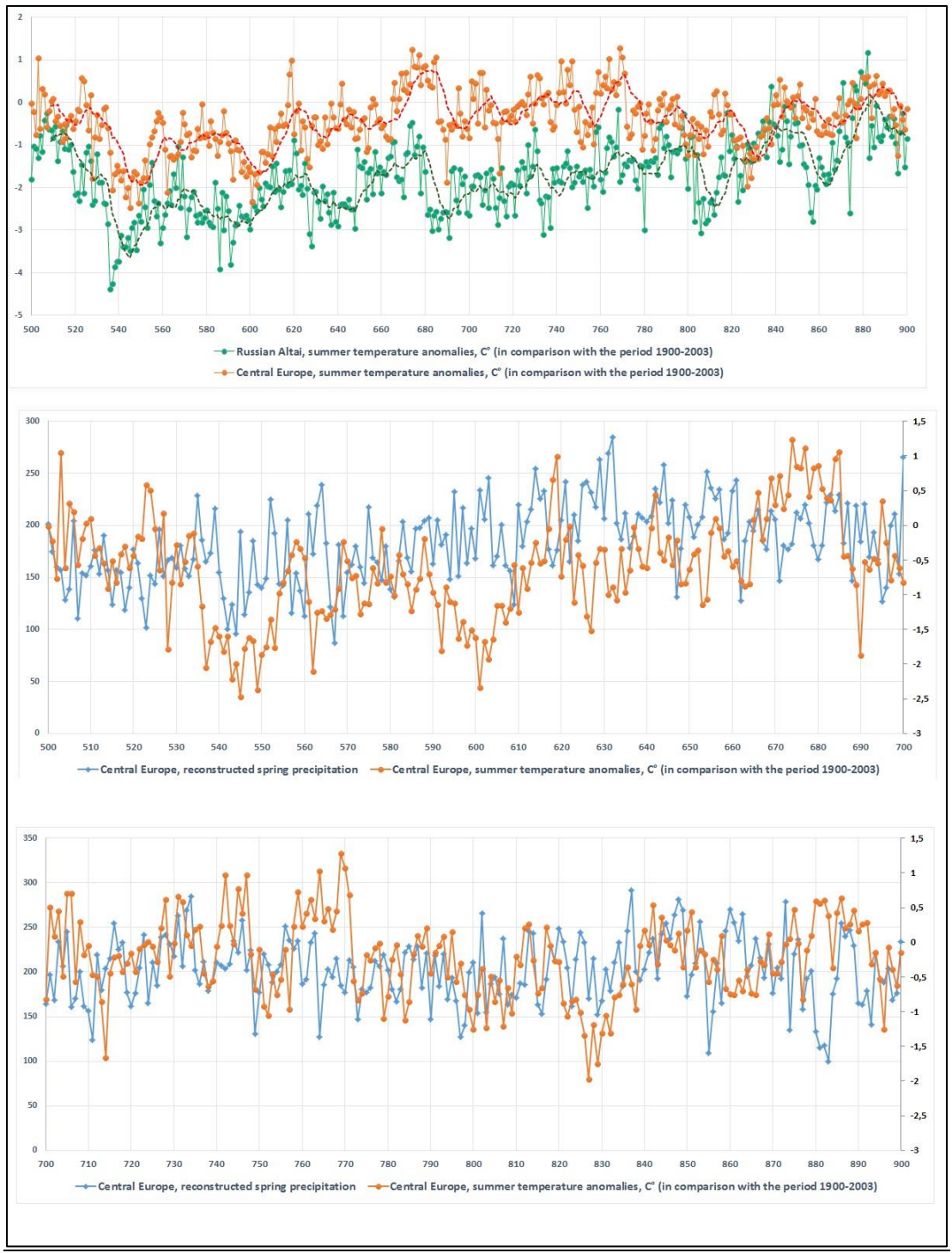

**Fig. 1 top**: Summer temperature anomalies in the Russian Altai (green) and in Central Europe (orange), C° (in comparison with the period 1900-2003), 500-900 AD; doted lines: 10years moving average; **middle**: summer temperature anomalies in Central Europe (orange), C° (in comparison with the period 1900-2003) and reconstructed spring precipitation in Central Europe (blue), 500-700 AD; **bottom**: summer temperature anomalies in Central Europe (orange), C° (in comparison with the period 1900-2003) and reconstructed spring precipitation in Central Europe (blue), 700-900 AD; (data: Büntgen et alii, Cooling and societal change; Büntgen et alii, Climate Variability; graph: Preiser-Kapeller, 2016).

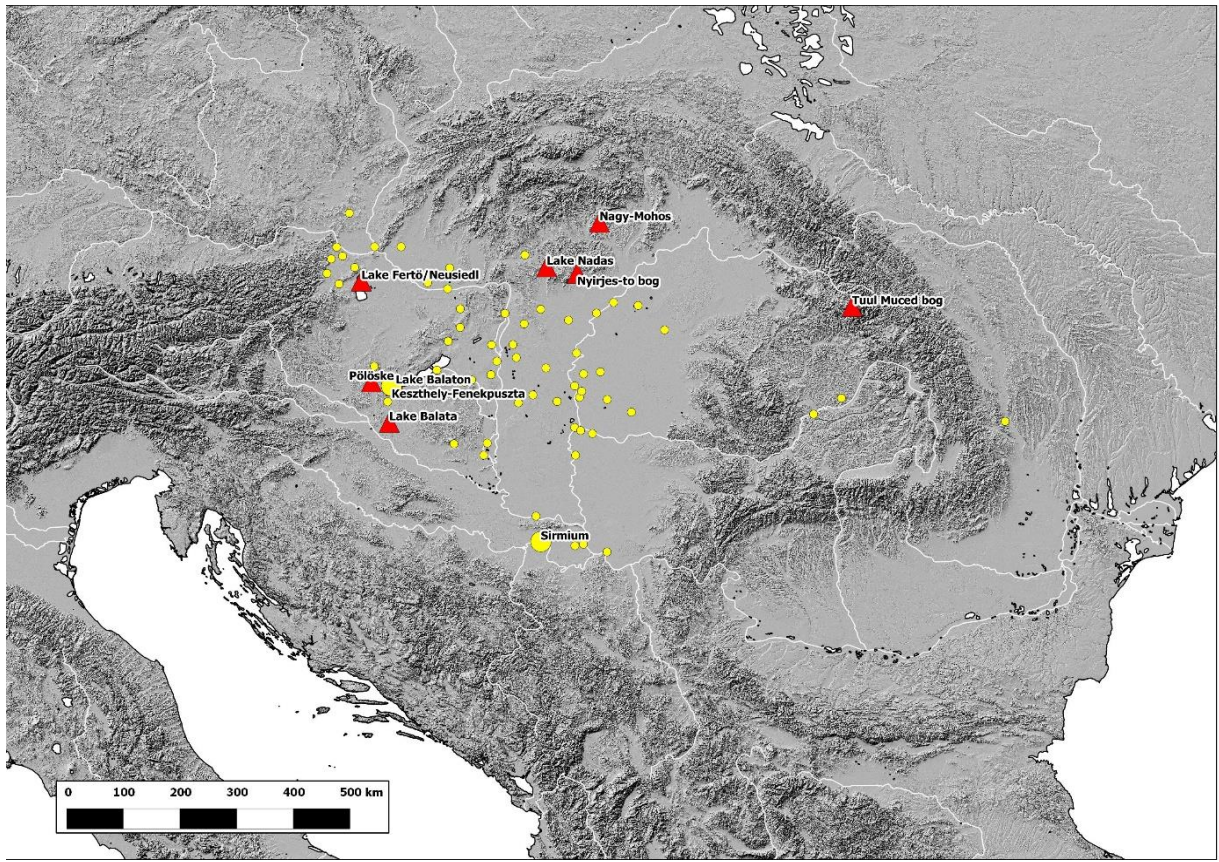

**Fig. 2** Map of selected sites from the Avar period (yellow circles) and sites of origin of palaeoenvironmental data (red triangles) mentioned in the paper (Preiser-Kapeller 2016)

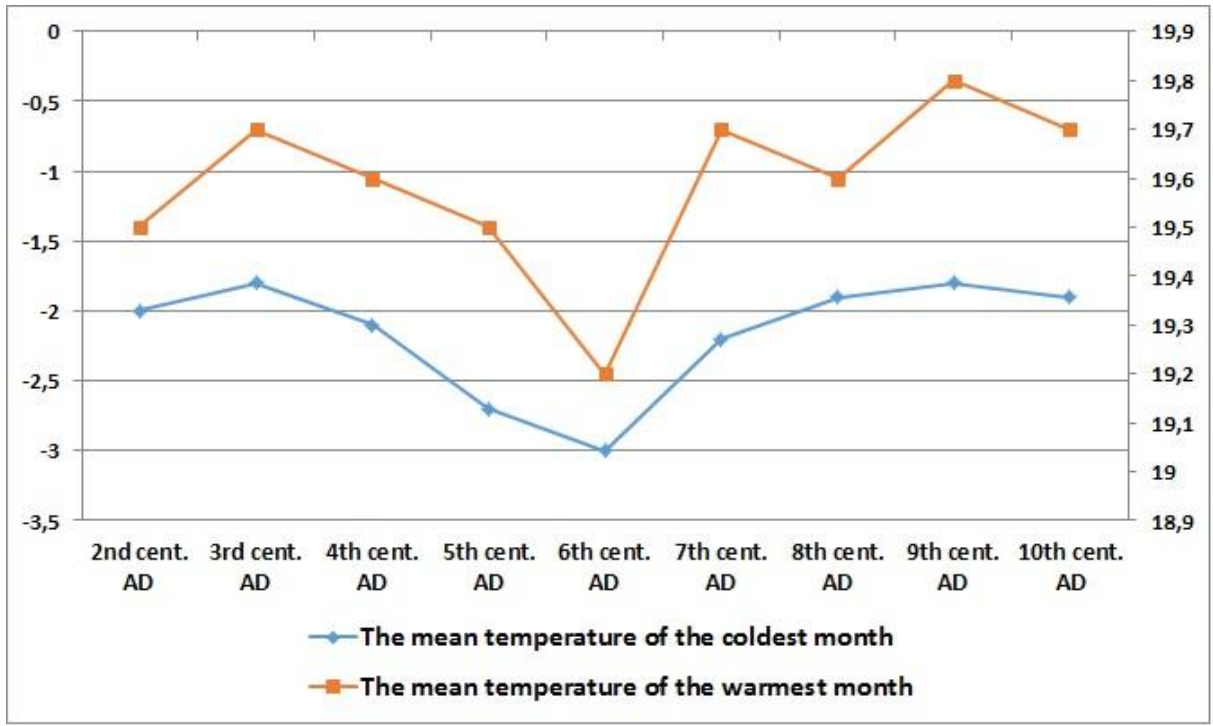

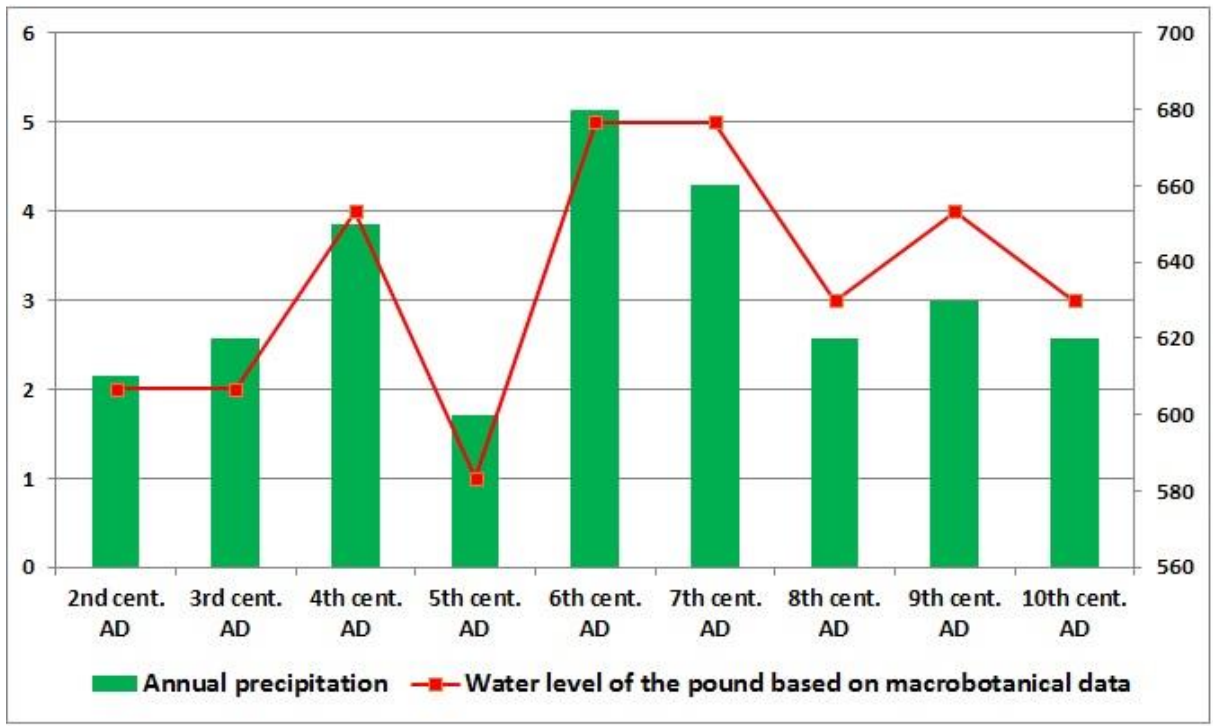

**Fig. 3 top**: Reconstructed mean temperatures of the coldest and the warmest month in the 2nd-10th cent. AD on the basis of data from Lake Nádas in northern Hungary; **bottom**: reconstructed mean annual precipitation and water level of the pound in the 2nd-10th cent. AD on the basis of data from Lake Nádas in northern Hungary (data: Sümegi et al., Middle Age; graph: Preiser-Kapeller, 2016)

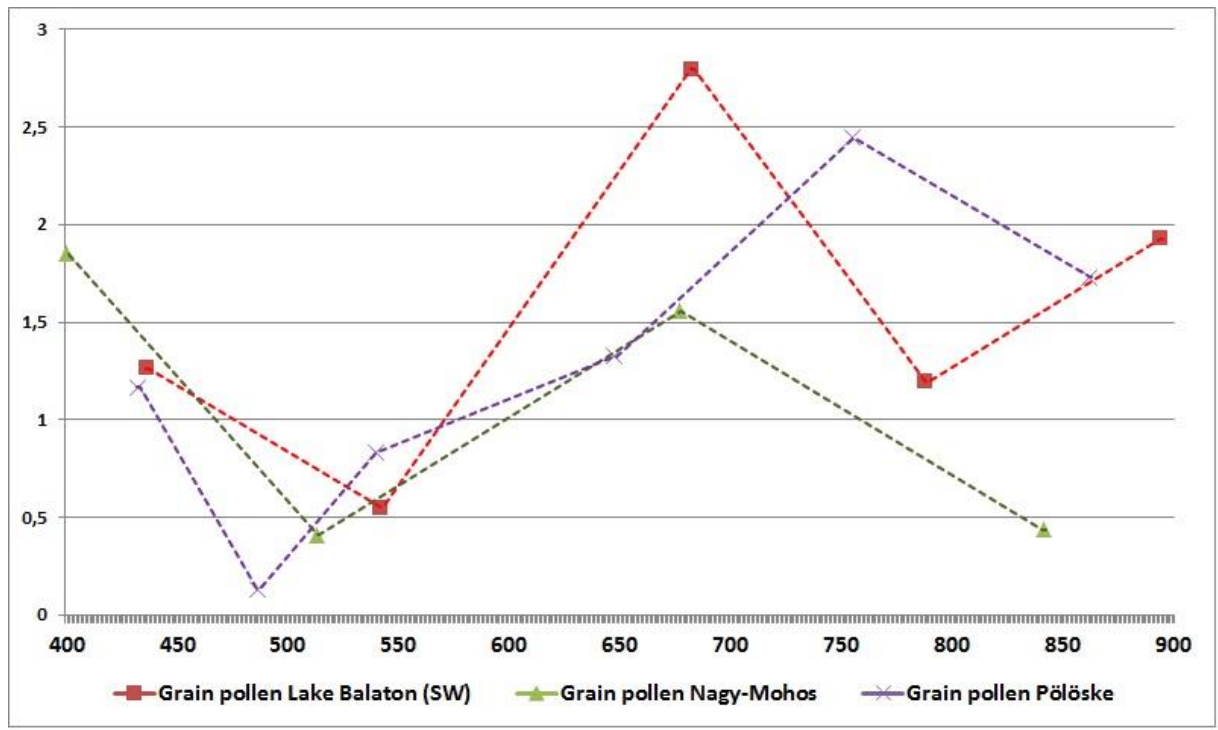

**Fig. 4** Indices of the concentration of grain pollen in samples from the southwest of Lake Balaton (red), Nagy-Mohos (green) and Pölöske (purple), 400-900 AD. Please note that the dating for these layers is highly problematic, which very much limits (or maybe even invalidates) their explanatory value (data: Juhász, Reconstitution palynologique; Zatykó / Juhász / Sümegi, Environmental archaeology; EPD: European Pollen Database [http://www.europeanpollendatabase.net/]; graph: Preiser-Kapeller 2016)

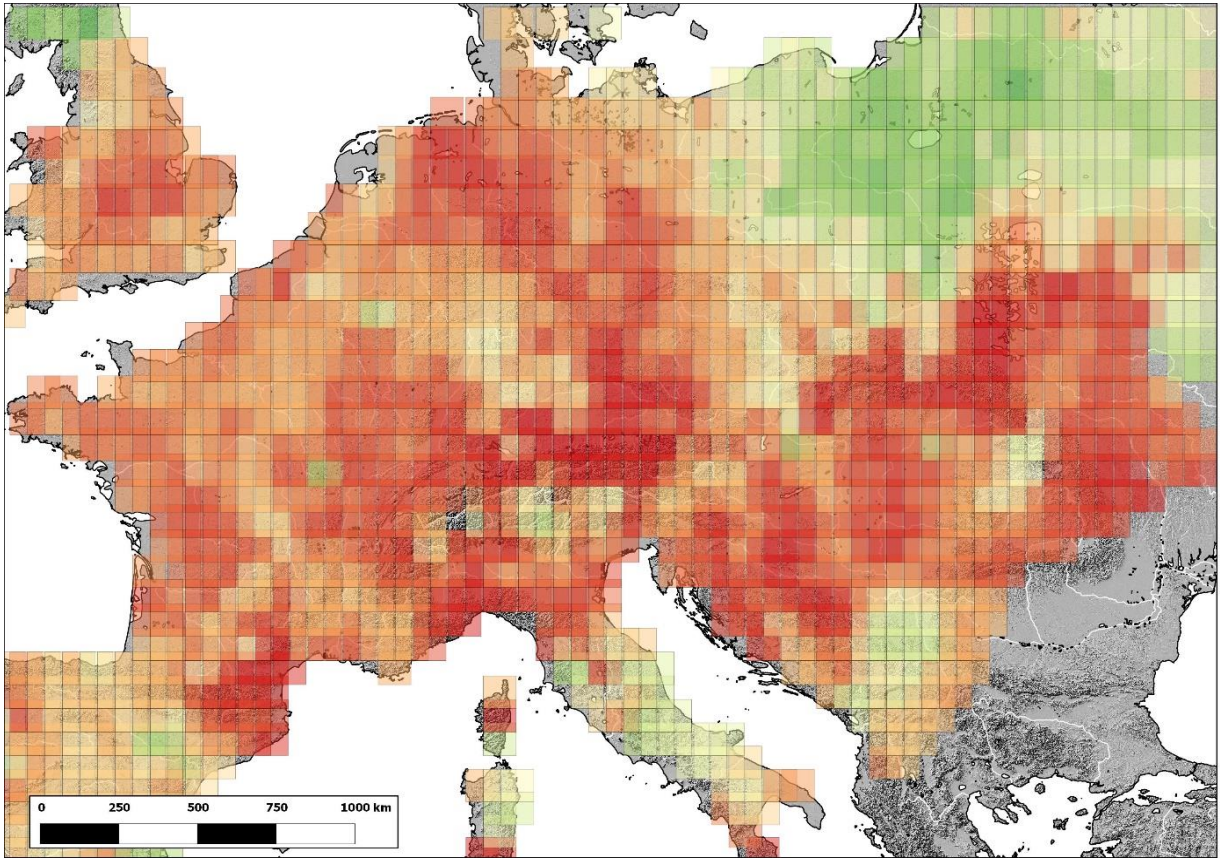

**Fig. 5** Reconstruction of summer wetness and dryness across central and western Europe for the year 810 AD; the colour scale from red to green shows the Palmer Drought Severity index, ranging from -4 or less (extreme drought) to +4 or above (extremely moist) (data: Cook et alii, Old World megadroughts; map: Preiser-Kapeller 2016)